\documentclass[aps,pra,twocolumn]{revtex4}   
\newcommand {\be}{\begin{equation}}
\newcommand {\ee}{\end{equation}}
\newcommand {\bey}{\begin{eqnarray}}
\newcommand {\eey}{\end{eqnarray}}
\usepackage{epsfig}
\usepackage{amsfonts}
\usepackage{amsmath}
\usepackage{mathbbol}

\begin{document}

\title{State space dimensionality in short memory hidden variable theories}

\author{Alberto Montina}
\affiliation{Perimeter Institute for Theoretical Physics, 31 Caroline 
Street North, Waterloo, Ontario N2L 2Y5, Canada}

\date{\today}

\begin{abstract}
Recently we have presented a hidden variable model of measurements 
for a qubit where the hidden variable state space dimension is 
one-half the quantum 
state manifold dimension. The absence of a short memory 
(Markov) dynamics is the price paid for this dimensional reduction. 
The conflict between having the Markov property and achieving
the dimensional reduction was proved in [A. Montina, Phys. Rev. A,
{\bf 77}, 022104 (2008)] using an additional hypothesis of 
trajectory relaxation. Here we analyze in more detail this
hypothesis introducing the concept of invertible process
and report a proof that makes clearer the role played
by the topology of the hidden variable space. This is accomplished
by requiring suitable properties of regularity of the conditional
probability governing the dynamics. In the case of minimal dimension 
the set of continuous hidden variables is identified with an object
living an $N$-dimensional Hilbert space, whose dynamics is described 
by the Schr\"odinger equation.
A method for generating the economical non-Markovian model for the 
qubit is also presented.
\end{abstract}
\maketitle

\section{Introduction}
One of the most peculiar features of quantum mechanics is the 
exponential growth of resources required to define the quantum state 
$|\psi\rangle$ of a composite system. It makes the direct
simulation of even a handful of particles impossible in practice.
This growth is due to the fact that 
$|\psi\rangle$ contains the full statistical information about 
the probabilities of any possible event, such as the joint probability 
$\rho(s_1,...,s_N)$ of obtaining the outcomes $\{s_1,...,s_N\}$
by measuring the $z$ axis components of $N$ $1/2$-spins. The information 
in $\rho(s_1,...,s_N)$ grows exponentially for a given accuracy
and considerably exceeds the classical information required to specify 
the actual measurement outcome. Since the quantum state is not a physically 
accessible observable, but can be statistically reconstructed only 
by performing many measurements on different replicas~\cite{noclone}, it 
is natural to wonder if this resource excess is strictly necessary 
to describe the actual state of a single realization. The quantum probabilities 
could be reproduced by a hidden variable theory where a single system 
carries less information than the quantum state. 
In such a theory the quantum state $|\psi\rangle$ is
mapped to a probability distribution on a space $X$ of hidden
variable states, i.e.,
\be\label{map1}
|\psi\rangle\rightarrow \rho(X|\psi).
\ee
It is clear that the sampling space $X$ can have in principle a smaller 
dimension than the quantum state manifold. For example the space 
of functions on a one-dimensional domain is infinite-dimensional 
and any finite-dimensional Hilbert space can be embedded within it.

In accordance
with recent terminology, we will refer to the actual state 
$X$ of a quantum system and the corresponding space as 
{\it ontic state} and {\it ontological space}~\cite{spekkens1},
respectively, and name the dimensional reduction of the
ontological space {\it ontological shrinking}. It is interesting 
to note that, in any known short memory (Markov) hidden 
variable theory, the dimension of the state space is never 
smaller than the quantum state manifold dimension.
As an example, in the de Broglie-Bohm model the wave-function
has the role of a field of physical quantities and is supplied
by additional variables describing the particle positions.

The ontological shrinking has a connection with the concept 
of classical "weak simulation"~\cite{jozsa,denNest} in quantum 
information theory. In a classical 
"strong simulation" of a quantum computer, the goal is to
evaluate the measurement probabilities with high
accuracy. This requirement is stronger than necessary,
since in a real quantum computer a single run does not
give the measurement probabilities and the output 
is a precise event.
The probabilities concern the behavior of many experimental 
realizations. The goal of a classical "weak simulation" is not 
to compute the probability weights, but the outcomes in accordance 
with the weights. There are examples of quantum circuits 
that cannot be efficiently simulated in the strong way,
but whose weak simulation is nevertheless tractable~\cite{denNest}.
In a hidden variable theory with reduced sampling space,
the evaluation of the actual dynamics of a single realization 
would require less resources than the computation of 
the quantum state dynamics.
Thus, the ontological shrinking could offer in a natural way 
an efficient method of "weak simulation" of quantum computers.

In the recent years
the possibility of a statistical representation of quantum
states on a reduced sampling space was discussed by various
authors~\cite{hardy,montina,monti2,brukner,galvao,monti3}.
The problem of the smallest dimension of the ontological space 
was posed in Ref.~\cite{montina}. It was subsequently proved that 
the ontological dimension cannot be smaller than the quantum 
state manifold dimension in the case of a Markov hidden variable 
theory with an additional hypothesis of trajectory
relaxation~\cite{monti2}. We will refer to this result
as the {\it no-shrinking theorem}. Recently we reported an 
example of hidden variable model of measurements for a qubit 
whose state space is one-dimensional, i.e., smaller than 
the two-dimensional Bloch sphere~\cite{monti3}.
As a consequence of the dimensional reduction, the dynamics
is not a Markov process. This counterexample makes evident that
the short memory hypothesis is strictly necessary for the proof
of the theorem in Ref.~\cite{monti2}.
In this article, we review the one-dimensional model providing 
a method for generating it and analyze in more detail the
hypothesis of trajectory relaxation. We define the concept of
{\it invertible process} and show that it is always possible to 
find a sub-region
of a compact ontological space where all the processes are
invertible. Discarding insignificant transient states and
considering only invertible processes, we present a new 
version of the {\it no-shrinking} theorem that makes clearer the 
role played by the topology of the hidden variable space. This
is accomplished by requiring reasonable properties of regularity 
of the conditional probability governing the dynamics and 
explicitly using them in the theorem proof. By the way,
it is useful to remind that the dimension is a topological 
property and is not defined by the space cardinality.

In section~\ref{sect1} we introduce the general framework
of a hidden variable theory. In section~\ref{sect2} the
economical ontological model in Ref.~\cite{monti3} is reviewed.
The properties of regularity and the concept of invertible process 
are introduced in section~\ref{theor_sect}, where we also prove
the no-shrinking theorem and discuss its consequences in terms of 
resource cost.
In the same section we show that in any Markov hidden
variable theory with minimal space dimension the set 
of continuous hidden variables can be identified with an object, 
living in an $N$-dimensional Hilbert space, whose dynamics is 
described by the Schr\"odinger equation.
In appendix, we report a systematic construction method to 
generate the model in Sec.~\ref{sect2}, starting from a particular 
form of the probability distribution.

\section{General framework}
\label{sect1}
In a general hidden variable theory the quantum state is translated 
into a classical language by replacing it with a probability
distribution on a sampling space $X$ of ontic states. We assume
that the ontological space is a $M$-dimensional manifold described
by an $M$-tuple $\vec x$ of continuous variables and a possible
discrete index $n$. The 
mapping~(\ref{map1}) is not the most general, since the probability 
distribution could depend on the preparation context. For example, a 
$1/2$-spin can be prepared in the {\it up} state by merely selecting 
the beam outgoing from a Stern-Gerlach apparatus or can be prepared 
in the state $|\uparrow\rangle+|\downarrow\rangle$ and then suitably 
rotated. In order
to account for this possible dependence, we add
suitable parameters $\eta$ that identify the preparation
context~\cite{spekkens2}, i.e.,
\be\label{state_map}
|\psi\rangle\rightarrow\rho(\vec x,n|\psi,\eta).
\ee
The set of ontological variables $X=(\vec x,n)$ contains the whole 
information about a single realization, thus the probability of any event
is conditioned only by it. In particular, for the measurement 
of the trace-one projector $|\phi\rangle\langle\phi|$, 
it is assumed that there exists a conditional probability $P$
for the event $|\phi\rangle$ given the ontic state $X$.
In general also $P$ could depend on the context of
the measurement, thus we introduce additional parameters
$\tau$ in the conditional probability, i.e.,
\be
|\phi\rangle\rightarrow P(\phi|\vec x,n,\tau).
\ee
The probability of the event $|\phi\rangle$ given 
$|\psi\rangle$ has to be equal to the quantum mechanical
probability,
\be\label{born}
\sum_n\int d^Mx P(\phi|\vec x,n,\tau)\rho(\vec x,n|\psi,\eta)=
|\langle\phi|\psi\rangle|^2.
\ee

Finally, it is assumed that the dynamics at the ontological
level is Markovian~\cite{gardiner}. 
The ontic state evolves from $X_1$ at time $t_1$ to 
$X_2$ at time $t_2$ according to a conditional probability 
$K(X_2,t_2|X_1,t_1)$.
$K$ satisfies the Chapman-Kolmogorov equation~\cite{gardiner} 
and is a delta distribution for $t_2=t_1$,
\be
K(\vec x_2,n_2,t_1|\vec x_1,n_1,t_1)=\delta_{n_1,n_2}\delta(\vec x_2-\vec x_1).
\ee
For a time-homogeneous process the transition probability depends
only on the time difference $t_2-t_1$,
\be\begin{array}{c}
K(X_2,t_2|X_1,t_1)=K(X_2,t_2-t_1|X_1,0) \\
\equiv K(X_2|X_1,t_2-t_1)
\end{array}
\ee

In order to link the quantum language with the classical one, 
we can label $K$ with the corresponding unitary operator 
$\hat U$. As for the state preparation and measurement, in general 
the conditional probability can depend on additional parameters 
$\chi$, the transformation context, i.e.,
\be
\hat U\rightarrow K_{\hat U,\chi}(X_2|X_1).
\ee
Indeed an operator $\hat U$ can be physically implemented in different
ways. For example, the spin rotation $e^{-i t\hat\sigma_x}$
can be performed directly rotating along the $x$ axis or implementing 
the three-step rotation $e^{-i\frac{\pi}4 \hat\sigma_y}
e^{-i t\hat\sigma_z} e^{i\frac{\pi}4 \hat\sigma_y}$ along
the $z$ and $y$ axes.
These two schemes are physically different and not
necessarily described by the same conditional probability.
By the way, it is useful to note that 
$K_{\hat U,\chi}(X|\bar X)$ for $\hat U={\mathbb 1}$ is
not necessarily the delta distribution $\delta(X-\bar X)$
for all the contexts.
For example the identity evolution can correspond to physically
performing the three step rotation  $e^{-i\frac{\pi}4 \hat\sigma_y}
e^{-i t\hat\sigma_z} e^{i\frac{\pi}4 \hat\sigma_y}$,
followed by the inverse transformation $e^{i t\hat\sigma_x}$.
The overall operation is not equal to doing nothing and
does not necessarily correspond to a delta peaked conditional
probability.

If the quantum state $|\psi\rangle$ evolves to
$\hat U|\psi\rangle\equiv|\bar\psi\rangle$, the associated
probability $\rho(X|\psi,\eta)$ evolves to
\be
\rho(X|\bar\psi,\bar\eta)\equiv\int dY K_{\hat U,\chi}
(X|Y)\rho(Y|\psi,\eta).
\ee
Some regularity properties of $K_{\hat U,\chi}$ will be discussed
in Sec.~\ref{theor_sect}.
Any short memory hidden variable theory has this general
structure.

\section{Economical ontological model of measurements}
\label{sect2}

In this section we show that if the Markov condition
is not assumed, then the ontological space can be
smaller than the quantum state manifold. This goal
is achieved by explicitly providing a one-dimensional
hidden variable model for a qubit~\cite{monti3}. 
Its systematic construction is discussed in the Appendix.

For the moment we introduce a model working only for a 
subset of preparation states. The extension to the
whole quantum state manifold will be discussed later on.
The ontological space is given by a
continuous real variable $x$ and a discrete index $n$ that
takes the two values $0$ and $1$. It is convenient
to represent the quantum state $|\psi\rangle$ and the event
$|\phi\rangle$  by means of the Bloch vectors
$\vec v\equiv \langle\psi|\vec\sigma|\psi\rangle$ and
$\vec w\equiv \langle\phi|\vec\sigma|\phi\rangle$,
where $\vec\sigma=(\hat\sigma_x,\hat\sigma_y,\hat\sigma_z)$,
$\hat\sigma_i$ being the Pauli matrices.

\begin{figure}[h!]
\epsfig{figure=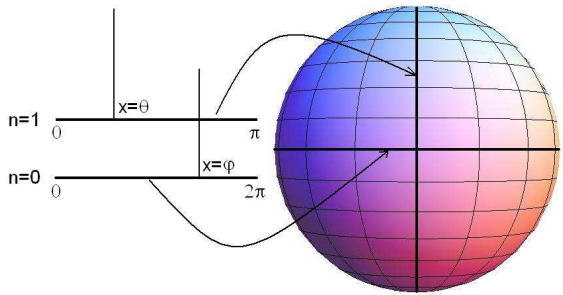,width=8.cm}
\caption{One-dimensional ontological space $\{x,n\}$ at left. 
According to Eq.~(\ref{distr_model}), each point of the Bloch 
sphere is associated with a probability distribution on
$\{x,n\}$ having two delta peaks.}
\label{fig1}
\end{figure}

The probability distribution associated with the state $\vec v$
is
\be\label{distr_model}
\rho(x,n|\vec v)=\sin\theta\delta_{n,0}\delta(x-\varphi)+
(1-\sin\theta)\delta_{n,1}\delta(x-\theta),
\ee
where $\theta$ and $\varphi$ are respectively the zenith and
azimuth angles in the spherical coordinate system
\be\begin{array}{l}
v_x=\sin\theta\cos\varphi,  \\ 
v_y=\sin\theta\sin\varphi,  \\
v_z=\cos\theta. 
\end{array}
\ee
Thus, when the quantum state $\vec v$
is prepared, the index $n$ takes the value $0$ or $1$ with
probability $\sin\theta$ or $1-\sin\theta$ and the continuous variable 
is equal to the zenith or azimuth angle according to the value 
of $n$ (see Fig.~\ref{fig1}). 

The conditional probability $P(\vec w|x,n)$ for an event $\vec w$ 
with $w_z>0$ is defined as follows 
\be
P(\vec w|x,0)=1+\frac{w_x \cos x+ w_y \sin x-\sqrt{1-w_z^2}}{2},
\ee
\be
P(\vec w|x,1)=1+\frac{\sqrt{1-w_z^2}\sin x+w_z\cos x-1}
{2(1-\sin x)}.
\ee
The events $\vec w$ with $w_z<0$ correspond simply to the non-occurrence
of the events $-\vec w$ with $w_z>0$, i.e, $P(-\vec w|x,n)=1-P(\vec w|x,n)$.

It is easy to prove that these probability functions fulfil
the condition (\ref{born}), that is,
\be
P(\vec w|\varphi,0)\sin\theta+P(\vec w|\theta,1)(1-\sin\theta)=
\frac{1}{2}\left(1+\vec w\cdot\vec v\right)
\ee

As shown in Ref.~\cite{monti3}, the conditional probabilities 
$P(\vec w|x,0)$ and $P(\vec w|x,1)$ are 
always smaller than or equal to $1$, but $P(\vec w|x,1)$ is positive
only if $\theta<\theta_0\equiv\arccos(\frac{3}{5})\simeq 53.13$ 
degrees.

Thus, the model in this form works only for a set of prepared
states whose Bloch vector lies inside a cone with aperture 
$2\theta_0$, the $z$ axis being the symmetry axis.
Since the positive region has non-zero measure in the
quantum state manifold, it is possible to extend the
model to the whole Bloch sphere by covering the manifold with 
a sufficiently large number of patch regions with different
coordinate systems and enriching
the ontic state with a finite quantity of information.
This was accomplished in Ref.~\cite{monti3} by adding a discrete 
index $m$ taking $12$ possible values, labelling $12$ different
regions of the Bloch sphere. 
See the referred paper for further details. 

This is a concrete example of ontological shrinking,
where the ontic state space is smaller than the
quantum state manifold, which is in this case the
two-dimensional Bloch sphere. 
It is a remarkable fact that a single realization of the 
ontic state $\{x,n\}$ contains less information than the 
quantum state. The whole information on $|\psi\rangle$
is contained in the probability distribution $\rho$.

The absence of a 
short memory description of the dynamics is the price
paid for the dimensional reduction. Indeed 
it is impossible to associate a positive 
conditional probability with each unitary evolution,
that is, it is impossible to satisfy the identity
\be\label{lin_evol}
\rho(x,n|\hat U\psi)=\sum_{\bar n}\int d\bar x
K_{\hat U}(x,n|\bar x,\bar n)
\rho(\bar x,\bar n|\psi)
\ee
with the probability distribution in Eq.~(\ref{distr_model}),
apart from the unitary evolution $e^{i t\hat\sigma_3}$.
Indeed in the other cases the dynamical equation of 
$\rho(x,n|\psi)$ is in general nonlinear.
Let us consider the unitary evolution with the Pauli
matrix $\hat\sigma_y$ as generator. The dynamical equations
of the Bloch vector are
\be\label{evol_v}
\begin{array}{l}
\frac{\partial v_x}{\partial t}=v_z,  \\
\frac{\partial v_y}{\partial t}=0,   \\
\frac{\partial v_z}{\partial t}=-v_x,
\end{array}
\ee
which correspond in spherical coordinates to
\be\begin{array}{l}
\label{der_sc}
\frac{\partial\varphi}{\partial t}=-\cot\theta\sin\varphi  \\
\frac{\partial\theta}{\partial t}=\cos\varphi.
\end{array}
\ee
Let $K(x,n|\bar x,\bar n,t)$
be the transition probability associated with the evolution
in Eq.~(\ref{evol_v}), we have from Eq.~(\ref{lin_evol}) that
\be
\frac{\partial\rho(x,n,t)}{\partial t}=
\sum_{\bar n}\int d\bar x \frac{\partial K}{\partial t}(x,n|\bar x,\bar n,t)
\rho(\bar x,\bar n),
\ee
that becomes by means of Eq.~(\ref{distr_model})
\be\begin{array}{l}
\sin\theta\delta_{n,0}\frac{\partial\delta(x-\varphi)}{\partial\varphi}
\frac{\partial\varphi}{\partial t}+ 
(1-\sin\theta)\delta_{n,1}\frac{\delta(x-\theta)}{\partial\theta}
\frac{\partial\theta}{\partial t}+ \\
\cos\theta\left[\delta_{n,0}\delta(x-\varphi)-\delta_{n,1}\delta(x-\theta)
\right] \frac{\partial\theta}{\partial t} =  \\
\frac{\partial K}{\partial t}(x,n|\varphi,0,t)\sin\theta+
\frac{\partial K}{\partial t}(x,n|\theta,1,t)(1-\sin\theta).
\end{array}
\ee
In particular, for $n=0$, and using Eq.~(\ref{der_sc}),
\be\begin{array}{l}
-\cos\theta\frac{\partial\delta(x-\varphi)}{\partial\varphi}
\sin\varphi+\cos\theta\delta(x-\varphi)
\cos\varphi= \\
\frac{\partial K}{\partial t}(x,0|\varphi,0,t)\sin\theta+
\frac{\partial K}{\partial t}(x,0|\theta,1,t)(1-\sin\theta).
\end{array}
\ee
Dividing both sides by $\sin\theta$ and differentiating
with respect to $\theta$, we obtain that
\be\begin{array}{c}
\frac{\partial}{\partial\theta}\left[
-\cot\theta\frac{\partial\delta(x-\varphi)}{\partial\varphi}
\sin\varphi+\cot\theta\delta(x-\varphi)
\cos\varphi\right]=  \\
\frac{\partial}{\partial\theta}
\left[\frac{\partial K}{\partial t}(x,0|\theta,1,t)
\frac{1-\sin\theta}{\sin\theta}\right],
\end{array}
\ee
This equation is not satisfied by any function $K$,
since the left-hand side is a function of both
$\theta$ and $\varphi$, whereas the right-hand side
depends only on $\theta$.
Thus, the dynamical equation of the probability
distribution~(\ref{distr_model}) is non-Markovian.

\section{Ontological shrinking and Markov processes}
\label{theor_sect}

In this section we will prove that the ontological
shrinking is in contradiction with a Markov
dynamics. This purpose is achieved by 
means of a very reasonable hypothesis concerning
the support of the conditional probability
$K(X_2|X_1,t)$. It will be introduced
in the following subsection.  
In subsection~\ref{subsec_S} we will introduce the
concept of invertible process and will show that
it is always possible to find a sub-region of a
compact ontological space where all the processes are
invertible. Then we prove the no-shrinking theorem in 
subsection~\ref{noshrinking}. In 
subsection~\ref{schrodinger}, it is shown that in
the case of minimal ontological dimension it is
possible to identify the set of continuous ontological
variables with a vector of the Hilbert space whose 
dynamics is given by the Schr\"odinger equation. In the 
last subsection we discuss the consequences of the 
no-shrinking theorem in terms of resource cost.

\subsection{Conditional probabilities and associated unitary
operators}
\label{trans_markov}

For the sake of simplicity, from now on we will omit without
loss of generality the discrete index $n$ in the definition of
the ontic state and assume that the ontological space is a
differentiable manifold whose points are identified by the 
vector $\vec x$. When the local Euclidean structure is not 
required, we will use the more generic symbol $X$ to indicate 
the ontic state.

In order that a transition probability describing
a Markov process makes physically sense, it has
to satisfy some conditions of regularity.
For our purpose, it is sufficient to require a very
weak condition. As a reasonable hypothesis, we assume that there 
exists a subset of the support of the conditional probability 
$K(X_2|X_1,t)$ that changes smoothly with respect to the condition 
$X_1$ and the evolution time $t$. Let us refine this statement in a 
more precise way. 

{\bf Property 1:} 
given a time-homogeneous process $X_1\rightarrow X_2$
with non-zero probability $K(\vec x_2|\vec x_1,t)$,
there exist an $M\times M$ matrix $\hat\lambda$
and a vector $\vec\alpha$ 
such that $K(\vec x_2+\hat\lambda\delta\vec x_1
+\vec\alpha\delta t|\vec x_1+\delta\vec x_1,t+\delta t)\ne0$ 
for any infinitesimal variation $\delta\vec x_1$ and $\delta t$.
The matrix $\hat\lambda$ and the vector $\vec\alpha$ 
are functions of $t$ and the process $X_1\rightarrow X_2$.

It is important to note that this property is fulfilled
for the very large class of Markov processes that involve
drift, diffusion and jumps (they are discussed
for example in Ref.~\cite{gardiner}). Furthermore, the
matrix $\hat\lambda$ and the vector $\vec\alpha$ are not
necessarily unique. For example in the case of a stochastic
process any choice of $\hat\lambda$ and $\vec\alpha$ is suitable,
since the conditional probability is a multidimensional smooth 
function whose support is the whole manifold.
For a pure deterministic
process both $\hat\lambda$ and $\vec\alpha$ are unique. The latter
gives the drift velocity of the ontic state. In the case 
of pure jumps with finite transition probability, one choice is 
$\vec\alpha=0$ and is unique if the jump distance cannot be arbitrarily 
small. One can imagine more complicated cases that involve
diffusion in some direction, drift in the other
ones and jumps, however also in these situations the
stated property is fulfilled for some $\hat\lambda$ and
$\vec\alpha$.

Property 1 strictly depends on the topology of the ontological space 
since it involves its local Euclidean structure. Using this structure
in the proof of the no-shrinking theorem is fundamental because the 
space dimension is not a property defined merely by the cardinality 
of the space. 

The quantum unitary evolution is a continuous process and 
corresponds to a trajectory in a Lie group manifold with the 
Hamiltonian as generator. 
For a time-homogeneous process, the 
unitary operator of the evolution has the form $e^{-i t \hat H}$, 
where $\hat H$ is the transformation generator and $t$ is the 
evolution time.  In general, it is not possible to directly
implement every generator of the $su(N)$ algebra by means of
a purely time-homogeneous process. In practice, only a small
set of evolutions (building blocks) can be directly generated
by a physical process. The other evolutions are obtained by 
suitably concatenating the building blocks. Given any generator 
$\hat G$, it will be possible in principle to experimentally
implement a unitary evolution $\hat U_1$, an physically
attainable time-homogeneous process $e^{-i t\hat H}$ and another
unitary evolution $\hat U_2$ such that 
\be\label{general_gen}
\hat U_2 e^{-i t\hat H}\hat U_1=e^{-i t\hat G}.
\ee
The operators $\hat U_i$ can be constructed by suitably 
concatenating physically attainable processes.
This allows us to associate with any unitary operator
$e^{-i t\hat G}$ a conditional probability $K(X_2|X_1,t)$
satisfying Property~1. Note that $t$ in $K$ does not 
correspond in general to the evolution
time of a purely time-homogeneous process. It is the
evolution time of only a part of the overall process
involving also the transformations $\hat U_i$.
This implies that for $t=0$ the conditional probability
$K(X_2|X_1,t)$ is not necessarily a delta distribution.
Furthermore, we can add a shift time $t_0$ to $t$ and
absorb the extra-term $e^{-i t_0\hat H}$ in 
Eq.~(\ref{general_gen}) into the operators $\hat U_1$
and $\hat U_2$. Thus, it is neither necessary to require
that $t$ is a positive quantity, as in the case of stochastic 
Markov processes where the propagation kernels are elements
of a semi-group. For our purpose, it sufficient that
$K(X_2|X_1,t)$ is defined in a neighborhood of $t=0$.

Let $\{\hat G_i\}$ with $i=1,...,D$ be a set of $D\equiv N^2-1$ 
generators of the Lie algebra. The unitary operators 
$e^{-i t \hat G_i}$ are associated with the conditional
probabilities $K_i(X_2|X_1,t)$. 

Any unitary evolution $\hat U$ in a region around the identity
can be constructed in the following way,
\be\label{chain}
\hat U(t_1,..,t_D)=e^{-i t_1 \hat G_1}...e^{-i t_D \hat G_D}=
\prod_{i=1}^D e^{-i t_i \hat G_i},
\ee
where
the variables $t_i$ parametrize the $SU(N)$ manifold. 

Let the linear operator
$$
\rho(X)\rightarrow\int dY K_i(X|Y,t)\rho(Y)
$$
be denoted by $K_i(t)$. The overall evolution $\hat U(\vec t)$ 
is associated with the conditional probability
\be\label{evol_kern}
\begin{array}{l}
K(X_2|X_1,t_1,..,t_D)=\left[ 
\prod_{i=1}^D K_i(t_i)\right](X_2|X_1),
\end{array}
\ee
the product order is such that the sum index grows from left 
to right. 

It is easy to prove a general property of the conditional
probability $K(X_2|X_1,\vec t)$. \newline
{\bf Property 2}: 
suppose that $K(\vec x_2|\vec x_1,\vec t)$ is different 
from zero for some process $X_1\rightarrow X_2$, then
there exists 
a $M\times D$ matrix $\hat\eta$ such that, for any small 
variation of the time parameters
$$
\delta\vec t=
\left(\begin{array}{c}\delta t_1\\ \delta t_2 
\\ ...\\ \delta t_D
\end{array}\right),
$$
the conditional probability $K(\vec x_2+
\hat\eta\delta\vec t|\vec x_1,\vec t+\delta \vec t)$ 
is different from zero. In the following, $\hat\eta$ will be
called the {\it shift matrix} of $K(X_2|X_1,\vec t)$. As with 
the matrix $\hat\lambda$ and the vector $\vec\alpha$, it is a 
function of $t$ and the process $X_1\rightarrow X_2$.

Proof: first, we consider the concatenation $K_{12}$ of two 
conditional probabilities, that is,
\be
K_{12}(X_2|X_1,t_1,t_2)\equiv\int dZ K_1(X_2|Z,t_1)
K_2(Z|X_1,t_2).
\ee
Let $X_1\rightarrow X_2$ be a process with non-zero 
transition probability $K_{12}(X_2|X_1,t_1,t_2)$. 
There exists a state $Z$ such that $K_1(X_2|Z,t_1)$ and 
$K_2(Z|X_1,t_2)$ are different from zero.
Property~1 implies that there exist
two vectors $\vec\alpha_1$ and $\vec\alpha_2$ and
a matrix $\hat\lambda_1$ such that
$$
K_1(\vec y+\hat\lambda_1\vec\alpha_2\delta t_2+
\vec\alpha_1\delta t_1|
\vec z+\vec\alpha_2\delta t_2,t_1+\delta t_1)\ne0,
$$
$$ 
K_2(\vec z+\vec\alpha_2\delta t_2|\vec x,t_2+\delta t_2)\ne0.
$$
Thus, the conditional probability 
$K_{12}(\vec x_2+\hat\eta\delta\vec t|\vec x_1,t_1+
\delta t_1,t_2+\delta t_2)$ is 
different from zero, where the columns of the $M\times2$ matrix
$\hat\eta$ are $\vec\alpha_1$ and $\hat\lambda_1\vec\alpha_2$.
The property can be proved by induction for any concatenation of
processes $K_i$.
$\square$

There is an important direct consequence of this property.\newline
{\bf Lemma 1}:
A process $X_1\rightarrow X_2$ with non-zero probability
$K(X_2|X_1,\vec 0)$ is associated 
with a $D_s$-dimensional manifold of unitary transformations
$\hat U(\delta\vec t)$, where $D_s\ge D-M$, 
$M$ being the ontological space dimension. 
The manifold is identified by the $M$ equations
\be\label{constr_mani}
\sum_{j=1}^D\eta_{ij}\delta t_j=0, \text{ with }
i=1,...,M,
\ee
where $\hat\eta$ is the shift matrix of $K(X_2|X_1,\vec 0)$.
In particular, $D_s=D-N_I$, $N_I(\le M)$ being the number of
independent equations in the constraints~(\ref{constr_mani}).

Proof: because of Property~2, 
there exists a $M\times D$ matrix $\hat\eta$ such that
$K(\vec x_2+\hat\eta\delta\vec t|\vec x_1,\delta\vec t)$ 
is different from zero, i.e, $\vec x_1\rightarrow\vec x_2+
\hat\eta\delta\vec t$ is a process associated with 
the unitary evolutions $\hat U(\delta\vec t)$.
In particular the submanifold of unitary evolutions with
$\hat\eta\delta\vec t=0$ is associated with the
process $X_1\rightarrow X_2$. Its dimension $D_s$ is
equal to $D-N_I$, $N_I$ being the number of independent equations
in $\hat\eta\delta\vec t=0$. Since $N_I\le M$, we have
$D_s\ge D-M$.

The actual value of $D_s$ depends on the number $N_I$ of 
independent constraints in the vectorial equation 
$\hat\eta\delta\vec t=0$. 
For example, $D_s$ is equal to $N^2-1$ for $\hat\eta=0$, 
which is the case if $K_i(\vec y|\vec x,t)$ are $M$-dimensional 
diffusive processes.

\subsection{Set {\cal S}(X) and its symmetry property}
\label{subsec_S}
In an ontological model
the quantum state is associated with a probability distribution 
$\rho(X|\psi,\eta)$ according to the mapping~(\ref{state_map}).
It is useful to introduce the following definition
of the set ${\cal S}(X)$. \newline
{\bf Definition 1}: a vector $|\psi\rangle$ of the Hilbert
space is in ${\cal S}(X)$ if and only if there exists a 
context $\eta$ such that the probability $\rho(X|\psi,\eta)$ 
is different from zero for the state $X$.

In other words, the set ${\cal S}(X)$ contains every quantum 
state that is compatible with the occurrence of the ontic state 
$X$. As discussed in Ref.~\cite{monti2}, the set $\cal S$ cannot 
lose vectors along its evolution. More precisely, if
$X\overset{\hat U}{\rightarrow}Y$ is a non-zero probability 
process associated with the unitary evolution $\hat U$, then
$\hat U{\cal S}(X)\subseteq{\cal S}(Y)$. This is a direct
consequence of the definition. Indeed, if $|\psi\rangle$ 
is in ${\cal S}(X)$, then there exists a probability
distribution associated with $|\psi\rangle$ such that $X$ is 
in its support. Since $X\rightarrow Y$  is a non-zero
probability process, then $Y$ is in the support of a probability 
distribution associated with the evolved quantum state
$\hat U|\psi\rangle$, that is, 
$|\psi\rangle\in {\cal S}(X)\Rightarrow \hat U|\psi\rangle
\in {\cal S}(Y)$. The opposite implication $|\psi\rangle\in 
{\cal S}(X)\Leftarrow \hat U|\psi\rangle\in {\cal S}(Y)$ 
is not trivially satisfied. Thus, the set $\cal S$ cannot 
lose vectors, but in principle it could grow acquiring
vectors. 

The opposite implication can be deduced by assuming that each
process is invertible. \newline
{\bf Definition 2}: a non-zero probability process 
$X_1\overset{\hat U_1}{\rightarrow}X_2$ is said to be {\it invertible}
if there exists a unitary operator $\hat U_2$ such that
$X_2\overset{\hat U_2}{\rightarrow}X_1$ is a non-zero probability 
process. 
\newline
{\bf Property 3}: every process is invertible.

Note that the operator $\hat U_2$ associated with the inverse 
process is not required to be necessarily the inverse of $\hat U_1$.
Property~3 is very reasonable and is satisfied by any known 
hidden variable theory. Indeed a state connected to
other states by means of a non-invertible process 
would be only transient and could be safely eliminated by the 
theory. The fact that the transient states are insignificant
is made clearer if it is assumed that the ontological space
is compact. \newline
{\bf Proposition}: if the ontological space is compact, then
the processes in any series  
$X_1\overset{\hat U_1}\rightarrow X_2\overset{\hat U_2}
\rightarrow X_3\overset{\hat U_3}\rightarrow ...$
become closer and closer to being invertible.

Proof: a metric ontological space is compact if it is closed
and bounded.
Let $B_R(X)$ be $M$-dimensional balls with
radius $R$ and center $X$.
Suppose that for any $R$ there exists a series 
$X_1\overset{\hat U_1}\rightarrow X_2\overset{\hat U_2}
\rightarrow X_3\overset{\hat U_3}\rightarrow ...$ 
with an infinite number of processes that take a state $X$
away from its ball $B_R(X)$. Then there exists a subseries
$Y_1\rightarrow Y_2\rightarrow Y_3\rightarrow ...$ where
$Y_m\notin B_R(Y_n)$ for every $m>n$, but this is impossible
because the space is bounded. 
Along the series, the processes connecting an element to the 
following ones become in fact closer and closer to being invertible. 
$\square$

Discarding insignificant transient states and taking for granted Property~3, 
we can prove the second lemma.
\newline
{\bf Lemma 2}: Assuming Property 3, if
$X_1\overset{\hat U_1}{\rightarrow} X_2$ is a non-zero
probability process, then $\hat U_1{\cal S}(X_1)={\cal S}(X_2)$.

Proof: 
the process $X_1\overset{\hat U_1}{\rightarrow} X_2$ is 
allowed, thus
\be\label{incl_go}
\hat U_1 {\cal S}(X_1)\subseteq {\cal S}(X_2).
\ee
Since there exists an operator $\hat U_2$
such that $X_2\overset{\hat U_2}{\rightarrow}X_1$ is a non-zero
probability process, then we have also that
\be\label{incl_back}
\hat U_2 {\cal S}(X_2)\subseteq {\cal S}(X_1).
\ee
These two relations imply that
\be\label{incl1}
\hat U {\cal S}(X_2)\subseteq {\cal S}(X_2),
\ee
where $\hat U=\hat U_1\hat U_2$. By iteration we obtain
that
\be
\hat U^n {\cal S}(X_2)\subseteq {\cal S}(X_2),
\ee
for any integer $n$. For a finite dimensional Hilbert space, it 
is always possible to find an integer $n$ such that $\hat U^n$ is
very close to the inverse operator $\hat U^{-1}$. Thus,
$\hat U^{-1} {\cal S}(X_2)\subseteq {\cal S}(X_2)$,
that is,
\be\label{incl2}
\hat U{\cal S}(X_2)\supseteq {\cal S}(X_2).
\ee
From inclusions~(\ref{incl1},\ref{incl2}) we have that
\be\label{symm_set}
\hat U{\cal S}(X_2)={\cal S}(X_2).
\ee
Applying the operator $\hat U_1$ to both sides of 
inclusion~(\ref{incl_back}) and using Eq~(\ref{symm_set}),
the inclusion
\be
{\cal S}(X_2)\subseteq \hat U_1{\cal S}(X_1)
\ee
is obtained. This relation and inclusion~(\ref{incl_go}) imply that
\be\label{evol_S}
\hat U_1 {\cal S}(X_1)={\cal S}(X_2)
\ee
and the lemma is proved. $\square$ \newline
Note that, for symmetry reasons, a similar equation holds
also for the inverse process, that is,
\be
\hat U_2 {\cal S}(X_2)={\cal S}(X_1).
\ee

In Ref.~\cite{monti2} we proved the following property
for ${\cal S}(X)$. \newline
{\bf Lemma 3}: the set ${\cal S}(X)$ cannot contain
every vector of the Hilbert space. Equivalently, the set
${\cal S}(X)$ is not invariant with respect to the
group $SU(N)$.

Proof by contradiction:
suppose that ${\cal S}(X)$ contains
every vector of the Hilbert space, then it contains 
in particular also two orthogonal vectors. This means
that there exist two overlapping distributions
associated with two orthogonal quantum states. But
this is impossible because two orthogonal states
can be perfectly discriminated by a measurement
\cite{hardy,spekkens2}.
Indeed the probability of obtaining
$|\psi\rangle$ given $|\psi\rangle$ is 
\be
\int dX P(\psi|X,\tau)\rho(X|\psi,\eta)=
|\langle\psi|\psi\rangle|^2=1.
\ee
This implies that $P(\psi|X,\tau)$ is 
equal to $1$ in the support of $\rho(X|\psi,\eta)$.
However the probability of $|\psi\rangle$ given
an orthogonal state $|\psi_\perp\rangle$ is
\be
\int dX P(\psi|X,\tau)\rho(X|\psi_\perp,\bar\eta)=
|\langle\psi|\psi_\perp\rangle|^2=0.
\ee
This implies that $\rho(X|\psi_\perp,\bar\eta)$ cannot
be different from zero if $X$ is in the support
of $\rho(X|\psi,\eta)$, where $P(\psi|X,\tau)=1$.
$\square$

Finally, we enunciate the last lemma.\newline
{\bf Lemma 4}: Let $G$ be a Lie subgroup of the group $SU(N)$ 
acting on a $N$-dimensional Hilbert space $\cal H$ and
${\cal S}$ be a set of vectors in $\cal H$.
If the manifold dimension of $G$ is larger than $(N-1)^2$
and $\cal S$ is invariant with respect to $G$, then
$\cal S$ contains every vector of the Hilbert space, that 
is, $\cal S$ is invariant with respect to $SU(N)$.

Proof: 
any compact Lie group and their linear representations on
$\mathbb{C}^N$ are well-known. One can check that,
for $N\ne4$, the proper Lie subgroup of $SU(N)$ with largest 
manifold dimension is $SU(N-1)\times U(1)$.
Its dimension is equal to $(N-1)^2$. Thus, if the dimension
of $G$ is larger than $(N-1)^2$, then $G$ is in fact
$SU(N)$ and $\cal S$ contains every vector.
In the special case $N=4$, the symplectic
group $Sp(2)$ is the subgroup of $SU(4)$ with largest
dimension. A set of generators in the representation space
$\mathbb{C}^4$ is
\be
\hat \sigma_i^{(1)},\;\hat \sigma_i^{(1)}\hat\sigma_1^{(2)},\;
\hat \sigma_i^{(1)}\hat\sigma_2^{(2)},\;
\hat\sigma_3^{(2)},
\ee
with $i=1,2,3$, $\hat\sigma_i^{(k)}$ being two sets of
Pauli matrices acting on the tensor space 
$\mathbb{C}^2\otimes\mathbb{C}^2=\mathbb{C}^4$. It is
easy to check that these generators form the basis of
a Lie algebra with dimension equal to $10$, which is
larger than $(N-1)^2=9$. However also in this case
if the set ${\cal S}$ is invariant with respect to $Sp(2)$, 
then it contains every vector of the 
Hilbert space. Indeed, it is possible to show that
the only orbit of $Sp(2)$ is the whole Hilbert space,
the orbit of a subgroup being the set of states connected by
means of some subgroup element. It is sufficient to
prove that any vector is connected to
$|\uparrow\rangle|\uparrow\rangle$. A generic vector $|\psi\rangle$ 
has the form
\be
|\psi\rangle=\cos\theta|\uparrow\rangle|\phi_1>+
\sin\theta|\downarrow\rangle|\phi_2>,
\ee
where $|\phi_1\rangle$ and $|\phi_2\rangle$ are
two-dimensional vectors. This vector is connected
through the generators $\hat\sigma_i^{(1)}$ to a vector
with the form
\be
|\tilde\psi\rangle=\cos\theta|\uparrow\rangle|\uparrow>+
\sin\theta|\downarrow\rangle|\tilde\phi_2>.
\ee
Through the unitary operator 
$e^{\frac{\hat{\mathbb{1}}-\hat\sigma_3^{(1)}}{2}
\left(\theta_1\hat\sigma_1^{(2)}+\theta_2\hat\sigma_2^{(2)}
\right)}$ and a suitable choice of $\theta_i$, it is
possible to connect $|\tilde\psi\rangle$ to
\be
|\bar\psi\rangle=\cos\theta|\uparrow\rangle|\uparrow>+
e^{i\varphi}\sin\theta|\downarrow\rangle|\downarrow>.
\ee
This last vector is connect to $|\uparrow\rangle|\uparrow\rangle$
through the generators $\hat\sigma_3^{(1)}$ and 
$(\hat\sigma_1^{(1)}\hat\sigma_2^{(2)}+
\hat\sigma_2^{(1)}\hat\sigma_1^{(2)})/2=
-i\hat\sigma_+^{(1)}\hat\sigma_+^{(2)}+
i\hat\sigma_-^{(1)}\hat\sigma_-^{(2)}$, $\hat\sigma_\pm$
being the raising/lowering operators. $\square$

This theorem implies that the minimal number of parameters
required to define the orientation of a set $\cal S$ cannot
be smaller than $2N-2$, apart from the trivial case of
a completely symmetric set $\cal S$, which does not have
an orientation. This property was intuitively introduced in 
Ref.~\cite{monti2}, where we noted that a set with the
highest symmetry
$SU(N-1)\times U(1)$ requires $2N-2$ variables
to specify the orientation of its symmetry axis.

\subsection{No-shrinking theorem}
\label{noshrinking}
At this point we have sufficient tools to prove the
no-shrinking theorem. We will show that if the ontological
space dimension would be smaller than the quantum state
manifold dimension, then the set ${\cal S}(X)$
would be invariant with respect to the group $SU(N)$, in
contradiction with lemma~3.

{\bf Theorem}: 
If the dynamics in a Markov ontological theory satisfies 
Property~1 and all the processes are invertible, then
the ontological space dimension $M$ is not
smaller than the quantum state manifold dimension, that is,
$M\ge2N-2$, $N$ being the Hilbert space dimension.

Proof by contradiction: 
suppose that $M<2N-2$. Let us
consider the unitary evolution $\hat U(\vec t)$ defined
in Eq.~(\ref{chain}) and the associated conditional
probability $K$ given by Eq.~(\ref{evol_kern}).
We have from lemma~1 that a process $X\rightarrow Y$ 
with non-zero probability $K(Y|X,\vec 0)$
is associated with a 
$D_s$-dimensional manifold of unitary evolutions, say 
${\cal U}$, with 
\be\label{constr_Ds}
D_s\ge N^2-1-M>N^2-1-2N+2=(N-1)^2.
\ee
The manifold contains the identity.
As a consequence, the set ${\cal S}(X)$
evolves, for any unitary evolution in ${\cal U}$, to the same 
${\cal S}(Y)$. By means of lemma~2 we have that
\be
\hat U{\cal S}(X)={\cal S}(Y) ,\;\; 
\forall \hat U\in{\cal U}.
\ee
Since $\mathbb{1}\in{\cal U}$, 
\be
{\cal S}(X)={\cal S}(Y).
\ee
Thus,
\be
\hat U{\cal S}(X)={\cal S}(X) ,\;\; 
\forall \hat U\in{\cal U},
\ee
that is, ${\cal S}(X)$ is symmetric with respect 
to a group whose generators are algebraically
generated by $D_s$ generators. The manifold dimension of this 
group is equal to or larger than $D_s>(N-1)^2$. Because of
lemma~4, ${\cal S}(X)$ is invariant with respect to $SU(N)$,
but this is in contradiction with lemma~3. We conclude that
the ontological space dimension $M$ cannot be smaller than
the quantum state manifold dimension $2N-2$. $\square$

\subsection{Minimal dimension and Schr\"odinger equation}
\label{schrodinger}

It is possible to show that in the case of minimal 
ontological dimension the set of continuous variables
can be identified with a vector living in the Hilbert
space,
whose dynamics is described by the Schr\"odinger equation.
It does not necessarily coincide with the quantum 
state, but moves rigidly with respect to it.

Suppose that the ontological dimension is minimal, that
is, $M=2(N-1)$. From lemmas 1-4 we have that
the set ${\cal S}(X)$ has to be symmetric
with respect to a subgroup of $SU(N)$ with manifold
dimension equal to $N^2-1-M=(N-1)^2$. The subgroup
is $SU(N-1)\times U(1)$. Thus,
the set has a symmetry axis $|\phi\rangle$,
which is a vector in the Hilbert space and identifies
the orientation of $\cal S$. Since each ontic state $X$ is
associated with a set ${\cal S}(X)$, we have the mapping
\be\label{mapping_X_phi}
X\rightarrow |\phi\rangle.
\ee

Because of Lemma~2, in a process associated with the
unitary evolution $\hat U(t)$, the set $\cal S$ 
rotates rigidly to $\hat U(t){\cal S}$. In particular,
its symmetry axis satisfies the Schr\"odinger equation
\be\label{shrod}
i\frac{\partial|\phi\rangle}{\partial t}=
\hat H(t) |\phi\rangle,
\ee
where $\hat H(t)$ is the Hamiltonian that generates
$\hat U(t)$.

The mapping~(\ref{mapping_X_phi}) is surjective, but 
is not necessarily injective. However, since the number of 
continuous ontological variables is just sufficient to label
$|\phi\rangle$ up to a global phase, an additional
discrete index is sufficient to make the mapping
bijective,
\be\label{map_onto_state}
X\leftrightarrow (|\phi\rangle,n).
\ee

Thus, the ontic state in a minimal Markov theory 
is identified with a vector in the Hilbert
space and a possible additional discrete index. The
dynamics of the vector is given by the
Schr\"odinger equation. $|\phi\rangle$
is equal to the quantum state if the set ${\cal S}(X)$ 
contains only one state that coincides with the 
symmetry axis. This is the case in a wave-pilot
theory such as the de Broglie-Bohm mechanics.

It is interesting to note that requirements (\ref{shrod},
\ref{map_onto_state}) are
fulfilled by the Kochen-Specker model~\cite{kochen}, where the 
ontic state is identified by a Bloch vector having the same 
dynamical equation of the quantum state.

\subsection{Resource cost with round off error}

We have proved that in any hidden variable theory with 
short memory dynamics the dimension of the state space 
cannot be smaller than the quantum state manifold dimension 
$2N-2$. There is a relation between dimension and resource
cost required to identify an ontic state. By resource cost
we mean the quantity of information required to identify
an ontological state.
Obviously the information carried by an ontic state is
infinite, since the ontological space is continuous.
Thus, the resource cost as a function of the dimension
makes sense only in presence of a fixed round off error of
the continuous variables. 

One can find a scaling law between dimension and resource cost 
in the following way.
Suppose that the ontic state is identified by $M$ continuous
variables $x_i$, which are defined in a finite interval. With a 
suitable rescaling, we can assume that they run in the interval
between $0$ and $1$. The probability $P({\cal E}|\vec x)$ of an event 
$\cal E$ is conditioned by the ontic state $\vec x$. It is assumed
that $P({\cal E}|\vec x)$ is a smooth function of $\vec x$, but
the discussion could be extended to the case of a finite number of 
discontinuities. Let $g_i$ be the mean magnitude of 
$\left|\partial_{x_i} P({\cal E}|\vec x)\right|$.

Let each continuous variable be discretized by $n_i$ points.
This introduces a round off error $\Delta E$ in $P$ that scales as
$\sqrt{\sum_{i=1}^M (g_i/n_i)^2}$ for sufficiently large values of 
$n_i$. The number of bits required to identify the ontic state
on the lattice is proportional to the information 
\be
{\cal I}=\sum_{i=1}^M \log n_i.
\ee
For a fixed $\cal I$, using the Lagrange multiplier method we find
that the optimal choice of $n_i$ that gives the smallest error is
\be
n_i=\frac{g_i e^{\frac{\cal I}{M}}}{\bar g},
\ee
where $\bar g\equiv \left(\prod_i g_i\right)^{1/M}$ is the 
log-average of $g_i$. Thus, we have the scaling law
\be
\Delta E\sim\bar g M^\frac{1}{2} e^{-\frac{\cal I}{M}}.
\ee
The error exponentially decreases with ${\cal I}$ at a rate inversely 
proportional to $M$. For a fixed error, the information scales as
\be\label{scaling_law}
{\cal I}\sim M\log\frac{M^\frac{1}{2} \bar g}{\Delta E}
\ee

The no-shrinking theorem states that $M$ is not smaller than the 
quantum state manifold dimension. Thus, in a composite system, 
${\cal I}$ grows at least exponentially in the number of parts 
(for example the number of qubits in a quantum computer). 

It is possible to estimate a lower bound for $\bar g$.
Since the probability of an event goes from $0$ to $1$ and $x_i$
rambles about the interval $[0:1]$, it is reasonable to assume 
that $g_i$ cannot be smaller than a value around $1$, that is, 
$$
\bar g\gtrsim 1.
$$
Thus the linear growth of $\cal I$ with respect $M$ cannot be
mitigated by an exponential decrease of $\bar g$.

\section{conclusions}

We have shown that in any hidden variable theory with a 
short memory dynamics the ontological space dimension cannot be 
smaller than the quantum state manifold dimension. Thus, like
the quantum state, the ontic state necessarily carries for a
given accuracy an amount of information that grows exponentially 
in the number of subsystems.
In comparison with Ref.~\cite{monti2}, we have provided a better
justification of the relaxation hypothesis, showing that in a 
compact ontological space it is possible to find a trajectory along
which the system unavoidably converges towards
a region where all the processes are invertible.
Furthermore we have presented a hidden variable
model of measurement for a qubit whose ontological space
is one-dimensional, which is one-half the dimension of the
Bloch sphere. The corresponding dynamics is not Markovian, 
in accordance with the no-shrinking theorem.

This model provides a counterexample
making evident that the hypotheses of short 
memory is strictly necessary to prove the no-shrinking theorem. 
By dropping it, we have shown that a single realization can 
carry less information than the quantum state. More drastically,
we could drop the causality hypothesis. Indeed
the non-causality is implied also by the Bell
theorem and the Lorentz invariance of the hidden variable theory. 
Thus, there are two signs that
point to the same direction, that is, the rejection of causality at the
ontological level. The possibility of an ontological shrinking
for a general $N$-dimensional Hilbert space in a theory without
a Markov dynamics is an open question whose answer could provide 
a deeper understanding of the computational complexity in
quantum mechanics.

I wish to thank J. Wallman and R. W. Spekkens for the careful 
reading of the manuscript and useful suggestions.
Research at Perimeter Institute for Theoretical Physics is
supported in part by the Government of Canada through NSERC
and by the Province of Ontario through MRI.

\appendix

\section{Deriving the economical model for a qubit}
In this appendix we will present a systematic method to generate
the model reported in Sec.~\ref{sect1}. We will consider
the class of probability distributions with the form
\be\label{prob_distr}
\rho(x,n|\vec v)=r(n|\vec v)\delta\left[x-f_n(\vec v)\right],
\ee
where the pair $\left(x\in\mathbb{R},n\in\{0,1\}\right)$ is
the ontic state and 
$\vec v\equiv\langle\psi|\vec\sigma|\psi\rangle$ is the Bloch
vector corresponding to the quantum state $|\psi\rangle$.
The quantities $f_n(\vec v)$ are real functions.
The normalization of the distribution gives 
\be\label{norma}
r(0|\vec v)+r(1|\vec v)=1.
\ee
When a system is prepared in 
$\vec v$, at the ontological level there is a probability
$r(0|\vec v)$ [$r(1|\vec v)$] that the discrete index takes
the value $n=0$ ($n=1$). Correspondingly, the continuous variable
takes with certainty the value $f_0(\vec v)$ [$f_1(\vec v)$].
This class is the simplest one that fulfils some requirements.
As noted in Subsec.~\ref{subsec_S}, two orthogonal states
cannot have overlapping probability distributions, implying
that the support of $\rho(x,n|\vec v)$ cannot be the whole 
ontological space. This rules out the class of smooth analytical
distributions and leads us to consider the probability distributions
with zero-measure support as simplest case.
In particular, the distributions with a two-point
support are parametrized with $3$ real parameters [the positions 
$f_n(\vec v)$ of the points and the relative probability
weight $r(0|\vec v)-r(1|\vec v)$], which are sufficient
in order to cover the two-dimensional Bloch sphere.

For our purposes,
it is convenient to define the variables $x_n\equiv f_n(\vec v)$
and to use them to parametrize the quantum state. Thus, the
probability distribution becomes
\be\label{prob_distr2}
\rho(x,n|x_0,x_1)=r(n|x_0,x_1)\delta\left(x-x_n\right).
\ee
Let $|\phi\rangle$ be the event of some projective measurement.
As for the preparation state, we introduce the Bloch
vector $\vec w\equiv\langle\phi|\vec\sigma|\phi\rangle$ 
to label the event $|\phi\rangle$.
$P_n(\vec w|x)$ is the conditional probability of the event
$\vec w$ given the ontic state $(x,n)$. Equation~(\ref{born})
becomes
\be\label{born2}
\sum_{n=0}^1\int dx P_n(\vec w|x)\rho(x,n|\vec x)=
\frac{1+\vec w\cdot\vec v(\vec x)}{2}\equiv S,
\ee
where $\vec v(\vec x)$ gives
the Bloch vector $\vec v$ as a function of the parameters
$(x_0,x_1)\equiv\vec x$. The quantity 
$1/2(1+\vec w\cdot \vec v)$ 
is the Born probability
$|\langle\phi|\psi\rangle|^2$ in terms of the Bloch 
vectors.

Using Eq.~(\ref{prob_distr2}), Equation~(\ref{born2})
becomes after integration
\be\label{eq1_4}
P_0(\vec w|x_0) r(0|x_0,x_1)+P_1(\vec w|x_1) r(1|x_0,x_1)=
S.
\ee
Note that the conditional probability $P_n(\vec w|x_n)$ in
this equation depends only on the variable $x_n$. Thus,
in order to find it, we could consider the four
pairs $(x_0,x_1)$, $(x_0,y_1)$, $(y_0,x_1)$ and
$(y_0,y_1)$. Correspondingly we have four equations 
with the four unknown functions $P_0(\vec w|x_0)$,
$P_0(\vec w|y_0)$, $P_1(\vec w|x_1)$ and $P_1(\vec w|y_1)$.
These equations are the row elements of the vector equation
\be\label{matrix_eq}
\hat R\vec P=\vec S,
\ee
where 
\be
\hat R\equiv 
\left(
\begin{array}{cccc}
r(0|x_0,x_1) & 0 & r(1|x_0,x_1) & 0  \\
r(0|x_0,y_1) & 0 & 0  & r(1|x_0,y_1) \\
0 & r(0|y_0,x_1) & r(1|y_0,x_1) & 0  \\
0 & r(0|y_0,y_1) & 0 & r(1|y_0,y_1),
\end{array}
\right),
\ee
\be
\vec P\equiv
\left(
\begin{array}{c}
P_0(\vec w|x_0) \\
P_0(\vec w|y_0) \\
P_1(\vec w|x_1) \\
P_1(\vec w|y_1)
\end{array}
\right)
\ee
and
\be\label{source}
\vec S\equiv \frac{1}{2}
\left(
\begin{array}{c}
1+\vec w\cdot\vec v(x_0,x_1) \\
1+\vec w\cdot\vec v(x_0,y_1) \\
1+\vec w\cdot\vec v(y_0,x_1) \\
1+\vec w\cdot\vec v(y_0,y_1)
\end{array}
\right)
\ee

If $\hat R$ would be invertible, $P_n(\vec w|x_n)$ would be
a linear function of $\vec w$, but this is impossible. 
It can be proved by contradiction. Suppose that the conditional 
probabilities are linear functions of $\vec w$, that is, 
$P_n(\vec w|x_n)=C_n(x_n)+\vec d_n(x_n)\cdot\vec w$.
They have to be non-negative and smaller than or
equal to $1$. It is clear that $P_n(\vec w|x_n)$
are equal to $1$ at most for only one vector $\vec w$,
but this is in contradiction with the fact that the conditional
probabilities have to be equal to $1$ in the one-dimensional
manifold of vectors $\vec w$ such that $f_n(\vec w)=x_n$.
Thus, the equation
\be\label{det_equal0}
\det \hat R(x_0,x_1,y_0,y_1)=0
\ee
has to be satisfied for any $x_n$ and $y_n$. Its solution
is found by differentiating in $y_0$ and $y_1$,
\be
\left.\frac{\partial^2}{\partial y_0\partial y_1}
\det \hat R(x_0,x_1,y_0,y_1)
\right|_{y_0=x_0,y_1=x_1}=0,
\ee
that is,
\be\label{diff_cond}
r(0|x_0,x_1)^2 r(1|x_0,x_1)^2\partial_{x_0}\partial_{x_1} 
\log\frac{r(1|x_0,x_1)}{r(0|x_0,x_1)}=0.
\ee
The functions $r(n|\vec x)$ cannot be identically equal to
zero, thus the only acceptable solution that satisfies also
the constraint~(\ref{norma}) is
\bey\label{solu_R0}
r(0|x_0,x_1)=\frac{k_1(x_1)}{k_0(x_0)+k_1(x_1)} \\
\label{solu_R1}
r(1|x_0,x_1)=\frac{k_0(x_0)}{k_0(x_0)+k_1(x_1)},
\eey
where $k_i(x_i)$ are positive functions of the variable
$x_i$. It is easy to check that this is also a
solution of Eq.~(\ref{det_equal0}).
Thus, the fact that the conditional probabilities
$P_n(\vec w|x_n)$ cannot be linear in $\vec w$ 
allows us to find a constraint for the
probabilities $r(n|\vec x)$.

Since the determinant of $\hat R$ is equal to zero, there
is a constraint also for the $\vec S$ because of 
Eq.~(\ref{matrix_eq}). Let the row vector 
$\vec u^T$ be the left eigenvector of $\hat R$
with eigenvalue $0$, then we have from Eq.~(\ref{matrix_eq}) 
that
\be\label{source_constr}
\vec u^T\cdot\vec S=0.
\ee
By means of Eqs.~(\ref{solu_R0},\ref{solu_R1}),
after a bit of calculations we find that the left eigenvector
of $\hat R$ with zero eigenvalue is
\be\label{left_eig}
\vec u=\left(
\begin{array}{c}
k_0(x_0)^{-1}+k_1(x_1)^{-1} \\
-k_0(x_0)^{-1}-k_1(y_1)^{-1} \\
-k_0(y_0)^{-1}-k_1(x_1)^{-1} \\
k_0(y_0)^{-1}+k_1(y_1)^{-1}
\end{array}\right).
\ee
From Eqs.~(\ref{source},\ref{source_constr},\ref{left_eig})
we obtain by differentiation the condition
\be\begin{array}{c}
\left.\partial_{y_0,y_1}\left(\vec u^T\cdot\vec S\right)
\right|_{y_n=x_n}=0\Longrightarrow  \\
\left.\partial_{y_0,y_1}\left[\frac{1}{k_0(x_0)}+
\frac{1}{k_1(x_1)}\right]\vec v(x_0,x_1)\right|_{y_n=x_n}=0,
\end{array}
\ee
that is satisfied if
\be\label{eq_v}
\vec v(x_0,x_1)=\left[\frac{1}{k_0(x_0)}+\frac{1}{k_1(x_1)}\right]^{-1}
\left[\vec g_0(x_0)+\vec g_1(x_1)\right],
\ee
where $\vec g_0(x_0)$ and $\vec g_1(x_1)$ are generic vectorial 
functions.
This is the inverse of the equation $\vec x=\vec f(\vec v)$.
The functions $g_n(x_n)$ and $k_n(x_n)$ are constrained by the 
equation $\vec v^2=1$, that is,
\be\label{main_constr}
\left[\vec g_0(x_0)+\vec g_1(x_1)\right]^2=
\left[\frac{1}{k_0(x_0)}+\frac{1}{k_1(x_1)}\right]^2.
\ee
We will return to it later on.

From Eqs.~(\ref{eq1_4},\ref{solu_R0},\ref{solu_R1},\ref{eq_v})
and the identity $S=\left[1+\vec w\cdot\vec v(\vec x)\right]/2$
we find that
\be
\sum_{n=0}^1\left\{\left[P_n(\vec w|x_n)-\frac{1}{2}\right]k_n^{-1}(x_n)-
\frac{1}{2}\vec w\cdot\vec g_n\right\}=0.
\ee
Note that each term of the summation depends only on one of the variables
$x_0$ and $x_1$. Thus, the conditional probabilities
have the form
\bey\label{P0}
P_0(\vec w|x_0)=k_0(x_0)\left[\frac{1}{2}\vec w\cdot\vec g_0(x_0)+H(\vec w)\right]
+\frac{1}{2},   \\
\label{P1}
P_1(\vec w|x_1)=k_1(x_1)\left[\frac{1}{2}\vec w\cdot\vec g_1(x_1)-H(\vec w)\right]
+\frac{1}{2},
\eey
where $H(\vec w)$ is an additional function independent of $\vec x$. 
Since the probability $P_0(\vec w|x_0)$ has to be equal to $1$ for 
$\vec w=\vec v$, we find that 
\be\label{eq1_H}
H(\vec v)=\frac{1}{2}\left[\frac{1}{k_0(x_0)}-\vec v\cdot\vec g_0(x_0)\right].
\ee
where $x_0=f_0(\vec v)$. Similarly, from the condition $P_0(\vec v|x_0)=1$
we obtain the equation
\be\label{eq2_H}
H(\vec v)=-\frac{1}{2}\left[\frac{1}{k_1(x_1)}-\vec v\cdot\vec g_1(x_1)\right].
\ee
This last equation can be derived by 
Eqs.~(\ref{eq_v},\ref{eq1_H}) and the constraint $\vec v^2=1$. It is interesting 
to note that the model we are constructing works only for a subset of 
preparation states $\vec v$, as shown in Sec.~\ref{sect2},
thus the function $H(\vec w)$ is not necessarily given by
Eqs.~(\ref{eq1_H},\ref{eq2_H}) if $\vec w$ is outside that subset.

At this point we have almost everything, the last step is to find
the functions $k_n(x_n)$ and $\vec g_n(x_n)$ that solve
Eq.~(\ref{main_constr}). Differentiating this equation with
respect to $x_0$ and $x_1$, we have that
\be
\frac{\partial \vec g_0(x_0)}{\partial x_0}\cdot
\frac{\partial \vec g_1(x_1)}{\partial x_1}-
\left(\frac{\partial}{\partial x_0}\frac{1}{k_0(x_0)}\right)
\left(\frac{\partial}{\partial x_1}\frac{1}{k_1(x_1)}\right)=0,
\ee
that is, the Minkowski inner product between the two four-vectors
\be
\alpha(x_0)\equiv
\left(
\begin{array}{c}
\vec g_0(x_0)+\vec\gamma_0 \\ \frac{1}{k_0(x_0)}+\chi_0
\end{array}
\right);\;\;
\beta(x_1)\equiv
\left(
\begin{array}{c}
\vec g_1(x_1)+\vec\gamma_1 \\ \frac{1}{k_1(x_1)}+\chi_1
\end{array}
\right)
\ee
is equal to $0$, $\vec\gamma_n$ and $\chi_n$ being constant
vectors and scalars, respectively. Using the Einstein notation,
on the index contraction, the constraint is
\be\label{ortho}
\alpha_\mu(x_0)\beta^\mu(x_1)=0.
\ee

It is important to note that $\alpha$ and $\beta$ depend only on
one of the variables $x_0$ and $x_1$. 
Using Eq.~(\ref{ortho}), Equation~(\ref{main_constr}) becomes
$
(\vec g_0-\vec\gamma_1)^2-\left(\frac{1}{k_0}-\chi_1\right)^2
-(\vec\gamma_0+\vec\gamma_1)^2+(\chi_0+\chi_1)^2
=-(\vec g_1-\vec\gamma_0)^2+
\left(\frac{1}{k_1}-\chi_0\right)^2.
$
The left-hand and right-hand sides depend only on
$x_0$ and $x_1$, respectively, thus they have to be
equal to a constant $r_0$,
\be
\begin{array}{l}
(\vec  g_0-\vec\gamma_1)^2-\left(\frac{1}{k_0}-\chi_1\right)^2= 
(\vec\gamma_0+\vec\gamma_1)^2-(\chi_0+\chi_1)^2+r_0 \\
(\vec g_1-\vec\gamma_0)^2-
\left(\frac{1}{k_1}-\chi_0\right)^2=-r_0.
\end{array}
\ee
These equations and Eq.~(\ref{ortho}) are a convenient 
resettlement of Eq.~(\ref{main_constr}).

It is interesting to observe that the conditional probabilities
$P_n(\vec w|x_n)$ and Eqs.~(\ref{eq_v},\ref{main_constr}) are 
invariant with respect to the transformation 
\be\label{trans_inv}
\vec g_0\rightarrow\vec g_0+\vec t,  \\
\vec g_1\rightarrow\vec g_1-\vec t,
\ee
for a generic vector $\vec t$. This means that the constant
vectors $\vec\gamma_0$ and $\vec\gamma_1$ are redundant and
for example one could set $\vec\gamma_1=0$.

In order to solve Eq.~(\ref{ortho}), we have to consider
two possibilities: (a) one of the two four-vectors $\alpha$ and
$\beta$ spans a one-dimensional vectorial subspace and the 
other one lives in the orthogonal subspace; (b) the two
vectors span orthogonal two-dimensional spaces.
In the former case no solution exists such that $\vec v(x_0,x_1)$ 
is locally invertible (only a one-dimensional subspace of
the Bloch sphere is represented).
Thus, we consider the latter case. The most general solution
up to rotations, transformation~(\ref{trans_inv}) and
variable change $x_n\rightarrow F_n(x_n)$ is
\be\label{solu1}
\begin{array}{l}
\vec g_0(x_0)=
\left(\begin{array}{c}
\cos x_0 \\
\sin x_0 \sin\theta_0 \\
0
\end{array}\right), 
\frac{1}{k_0(x_0)}=\cos\theta_0\cos x_0+s\\
\vec g_1(x_1)=
\left(\begin{array}{c}
\cos \theta_0 \csc x_1 \\
0  \\
\cot x_1\sin\theta_0
\end{array}\right),
\frac{1}{k_1(x_1)}=\csc x_1-s,
\end{array}
\ee
where $\theta_0$ and $s$ are two free parameters. Only the former
is present in the mapping $\vec x\rightarrow\vec v(\vec x)$, given
by Eq.~(\ref{eq_v}). We have that
\be\label{orthog}
\vec v(x_0,x_1)=
\frac{1}{1+\cos\theta_0 u_x}
\left(\begin{array}{c}
\cos\theta_0+u_x \\
\sin \theta_0 u_y \\
\sin\theta_0 u_z
\end{array}\right),
\ee
where
\be\begin{array}{l}
u_x=\sin x_1\cos x_0,  \\ 
u_y=\sin x_1\sin x_0,  \\
u_z=\cos x_1. 
\end{array}.
\ee
The mapping is bijective by removing the poles at $x_1=0$ and
$x_1=\pi$,
\be\begin{array}{c}
0\le x_0\le 2\pi, \\
0< x_1<\pi.
\end{array}
\ee

\begin{figure}[h!]
\epsfig{figure=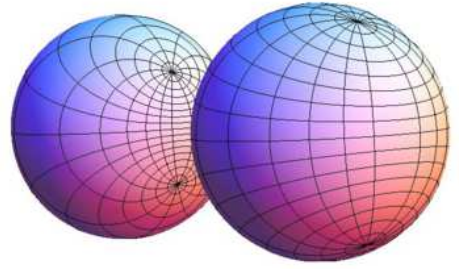,width=7.5cm}
\caption{Two orthogonal coordinate systems for $\theta_0=0.5\text{ rad}$
(left) and $\theta_0=1\text{ rad}$ (right).}
\label{fig2}
\end{figure}

The inverse of the vectorial function $\vec v(\vec x)$ in Eq.~(\ref{orthog}) 
gives the functions $f_n(\vec v)=x_n$, whose trigonometric functions are
\be\begin{array}{l}
\sin x_0=v_y \sin\theta_0/\Delta,   \\
\cos x_0=(v_x-\cos\theta_0)/\Delta, \\
\sin x_1=\frac{\Delta}{1-v_x \cos\theta_0}, \\
\cos x_1=\frac{v_z \sin\theta_0}{1-v_x\cos\theta_0},
\end{array}
\ee
with $\Delta\equiv \sqrt{(w_x-\cos\theta_0)^2+w_y^2\sin^2\theta_0}$.

Equation~(\ref{orthog}) provides a set of orthogonal coordinate systems of the 
sphere. The spherical coordinate system is obtained with $\theta_0=\pi/2$. 
Each system is mapped to another one by means of the M\"obius 
transformation~\cite{mobius}.
In Fig.~\ref{fig2} we report two coordinate systems for $\theta_0=0.5$, $1$. 
Both of them have two poles, but with different angular distance, that is
equal to $2\theta_0$. 

From Eqs.~(\ref{solu_R0},\ref{solu_R1},\ref{solu1}) we have that
\bey
& r(0|\vec x)=\frac{\sin x_1(s+\cos\theta_0 \cos x_0)} 
{1+\cos\theta_0 \cos x_0\sin x_1} & \\
& r(1|\vec x)=\frac{1-s \sin x_1}{1+\cos\theta_0\cos x_0\sin x_1}. &
\eey
They are positive if
\be
|\cos\theta_0|\le s\le 1.
\ee

$H(\vec w)$ is obtained by Eq.~(\ref{eq1_H}),
\be
H(\vec w)=\frac{s-\Delta}{2},
\ee
Finally, the conditional probabilities for the events are given
by Eqs.~(\ref{P0},\ref{P1}),
\be
P_0(\vec w|x)=1+\frac{(w_x-\cos\theta_0)\cos x+w_y\sin x\sin\theta_0-\Delta}
{2(s+\cos\theta_0\cos x)},
\ee
\be
P_1(\vec w|x)=1+\frac{(w_x\cos\theta_0-1)+w_z\cos x\sin\theta_0+\Delta \sin x}
{2(1-s \sin x)}.
\ee

The model in Sec.~\ref{sect2} is obtained for $\theta_0=\pi/2$ and $s=1$.


\begin{thebibliography}{20}
\bibitem{noclone} W. K. Wootters, W. H. Zurek, Nature {\bf 299}, 802 (1982).
\bibitem{spekkens1} R. W. Spekkens, Phys. Rev. A {\bf 75}, 032110 (2007).
\bibitem{jozsa} R. Jozsa, A. Miyake, Proc. R. Soc. A {\bf 464}, 3089 (2008).
\bibitem{denNest} M. Van den Nest, Quant. Inf. Comp. {\bf 10} 0258 (2010);
M. Van den Nest, arXiv:0911.1624.
\bibitem{hardy} L. Hardy, Stud. Hist. Phil. Sci. B {\bf 35}, 267 (2004).
\bibitem{montina} A. Montina, Phys. Rev. Lett. {\bf 97}, 180401 (2006);
A. Montina, J. Phys.: Conf. Ser. {\bf 67}, 012050 (2007).
\bibitem{monti2} A. Montina, Phys. Rev. A {\bf 77}, 022104 (2008).
\bibitem{brukner} B. Daki\'c, M. Suvakov, T. Paterek, and C. Brukner,
Phys. Rev. Lett. {\bf 101}, 190402 (2008).
\bibitem{galvao}  E. F. Galv\~ao, Phys. Rev. A {\bf 80}, 022106 (2009).
\bibitem{monti3} A. Montina, arXiv:1002.3139.
\bibitem{spekkens2} R. W. Spekkens, Phys. Rev. A {\bf 71}, 052108 (2005).
\bibitem{gardiner} C. W. Gardiner, "Handbook of Stochastic Methods"
(Springer, Berlin, 2004).
\bibitem{kochen} S. Kochen and E. P. Specker, J. Math. Mech. {\bf 17}, 59 (1967).
\bibitem{mobius} S. G. Krantz, "Geometric Function Theory: Explorations in 
Complex Analysis" (Birkh\"auser, Boston, 2006).



\end{thebibliography}
\end{document}